\documentstyle[12pt]{article}
\makeatletter
\@addtoreset{equation}{section}

\makeatother
\newcommand{\beq}{\begin{equation}}
\newcommand{\eeq}{\end{equation}}
\newcommand{\bea}{\begin{eqnarray}}
\newcommand{\eea}{\end{eqnarray}}
\newcommand{\half}{\frac{1}{2}}

\renewcommand{\a}{\alpha}
\renewcommand{\b}{\beta}
\newcommand{\e}{\epsilon}

\newcommand{\tB}{\tilde{B}}
\newcommand{\tC}{\tilde{C}}
\newcommand{\tR}{\tilde{R}}

\newcommand{\tX}{\tilde{X}}
\newcommand{\tY}{\tilde{Y}}
\newcommand{\tW}{\tilde{W}}
\newcommand{\smT}{\footnotesize T}
\newcommand{\dmu}{\partial_{\mu}}
\newcommand{\dmup}{\partial^{\mu}}
\newcommand{\derx}{{\bf\nabla}_x}
\newcommand{\dery}{{\bf\nabla}_y}
\newcommand{\derxa}{\partial_{a}}
\newcommand{\derxb}{\partial_{b}}
\newcommand{\derya}{\partial_{\a}}
\newcommand{\deryb}{\partial_{\b}}

\newcommand{\lnxal}{\ln(BC^{\bar{q}}X^{p_1-2}Y^{p_2}W^{p_t})}
\newcommand{\lnyal}{\ln(BC^{\bar{q}}X^{p_1}Y^{p_2-2}W^{p_t})}
\newcommand{\lnxallt}{\ln(\tB\tC^{\bar{q}}\tX^{p_1-2}\tY^{p_2}\tW^{p_t})}
\newcommand{\lnyallt}{\ln(\tB\tC^{\bar{q}}\tX^{p_1}\tY^{p_2-2}\tW^{p_t})}
\newcommand{\tdual}{\stackrel{\mbox{\smT}}{\longleftrightarrow}}
\newcommand{\tdua}{\mbox{\smT}\!\updownarrow}
\newcommand{\tBox}{\stackrel{\hat{}}{\Box}}
\begin{document}
\begin{titlepage}
\rightline{UM-TG-205}
\rightline{US-FT/1-98}
\rightline{hep-th/9801049}
\def\today{\ifcase\month\or
January\or February\or March\or April\or May\or June\or
July\or August\or September\or October\or November\or December\fi,
\number\year}
\vskip 8mm
\centerline{\Large \bf Rules for Localized Overlappings and}
\vskip 3mm
\centerline{\Large \bf Intersections of p-Branes}
\vskip 7mm
\centerline{\sc Jos\'e D. Edelstein$^{\,a,c}$,  
Liviu T\u{a}taru$^{\,b,}$\footnote{On leave of absence from
Dept. of Theoretical Physics, Univ. of Cluj, Romania.}
and Radu Tatar$^{\,b}$}
\vskip 7mm
\centerline{{\it $^a$ Depto. de F\'{\i}sica de Part\'{\i}culas,
Universidade de Santiago de Compostela}}
\centerline{{\it E-15706 Santiago de Compostela, Spain}}
\vskip 1mm
\centerline{{\it $^b$ Dept. of Physics, University of Miami,
Coral Gables, Florida 33146, USA}}
\vskip 1mm
\centerline{{\it $^c$ Depto. de F\'{\i}sica, Univ. N. de La Plata, CC 67, 
1900 La Plata, Argentina}}
\vskip 7mm
\centerline{\sc Abstract}
\vskip 0.13in

We determine an intersection rule for extremal p-branes which are
localized in their relative transverse coordinates by solving, 
in a purely bosonic context, the equations of motion of gravity 
coupled to a dilaton and n-form field strengths. The unique algebraic
rule we obtained does not lead to new solutions while it manages to 
collect, in a systematic way, most of the solutions (all those
compatible with our ansatz) that have appeared in the literature. 
We then consider bound states of zero binding energy where a third brane is
accomodated in the common and overall transverse directions. They are 
given in terms of non-harmonic functions. A different algebraic rule 
emerges for these last intersections, being identical to 
the intersection rule for p-branes which only depend on the overall
transverse coordinates. We clarify the origin of this coincidence.
The whole set of solutions in ten and eleven dimensional theories is 
connected by dualities and dimensional reductions. They are related to
brane configurations recently used to study non-perturbative phenomena
in supersymmetric gauge theories.

\end{titlepage}
\newpage
\setcounter{footnote}{0}

\section{Introduction}

In the last two years, there has been a growing amount of evidence in 
favour of the possible unification of the five known superstring
theories due to the existence of non-perturbative duality symmetries
\cite{HT,W}. An eleven dimensional theory has been conjectured, 
M-theory, that enables us to understand the duality properties of
string theory in a unified way \cite{To} and whose low-energy limit
is given by $D=11$ supergravity. A microscopic description of this theory 
has been proposed in terms of the large $N$ limit of a
supersymmetric quantum mechanical system of $N \times N$ matrices
\cite{BFSS}. It has been widely stressed in the literature
that one of the key ingredients that leads to the identification
of duality symmetries is given by a proper knowledge of the 
solitonic spectrum of these theories, this involving p-dimensional 
excitations called p-branes.

~

It is known that $D=11$ supergravity contains 
M2-branes and M5-branes, both playing a major r\^ole in 
the dynamics of M-theory. These branes preserve one half of the
supersymmetries, hence they are BPS states. In string theory, an
entire zoo of BPS p-brane solutions has been studied in the last 
few years. In type II theories, there are two kinds of p-branes
depending on the sector of the world-sheet theory where the
charge that they carry is originated. The NS-NS sector contains
the fundamental string and the solitonic fivebrane, the NS5-brane.
The p-branes that carry R-R charge have been shown
to be described by hypersurfaces where open strings
can end, called D-branes \cite{P}.

~

One of the most interesting aspects of this variety of branes
is given by the possibility of constructing composite brane configurations
--starting from the previously mentioned basic bricks-- that preserve
a certain amount of supersymmetries. This kind of configurations has
led very recently to a large number of celebrated results both in 
supergravity and supersymmetric gauge theories. Intersections of a large
number of D-branes has made possible to identify and count the
microscopic states corresponding --after compactification-- to certain
black hole geometries, in complete agreement with the semiclassical
entropy \cite{entro1,entro2,entro3}. 
Another remarkable fact is that some of the
non-perturbative properties of supersymmetric gauge theories in
various spacetime dimensions were found to have a natural explanation
in string theory, by studying the low-energy dynamics of a certain class 
of intersecting brane configurations \cite{HW,EGK,BSTY,BHOY,EJS,Witt}. 
In this scenario, it is of great interest to derive, on general grounds, 
a set of rules that states and classifies all possible brane 
configurations that preserve a specific amount of supersymmetries.

~

For the case of D-branes, by using the string theory representation of the
branes and duality, certain rules were derived in Refs.\cite{PCJ,S,To2}.
The case of intersecting M-branes was considered in \cite{PT} and
the study of intersecting p-branes starting from eleven dimensional
supergravity has led to the formulation of the so-called harmonic 
superposition rule \cite{Ts}. Intersection of both M-branes and
D-branes were subsequently classified in Ref.\cite{BREJS}. Another 
derivation of the intersection rules, not based on supersymmetry
arguments, was done by requiring that p-brane probes in q-brane
backgrounds feel no force and can thus create bound states with
vanishing binding energy \cite{Ts2}. 

~

More recently, a general rule determining how extremal branes can intersect
in a configuration with zero binding energy has been derived in 
Ref.\cite{AEH}\footnote{Also, following a slightly different approach, 
in Ref.\cite{ARV}.}.
This rule is obtained from the bosonic equations of motion of the 
low-energy theory and unifies in a remarkably simple way all classes
of branes in any spacetime dimension\footnote{It was also generalized in 
Ref.\cite{Oh} to include intersections of non-extreme p-branes.}.
The kind of configurations considered there are tranlational invariant in
all directions tangent to any participating brane. This is an appropriate
restriction if one is to consider toroidal compactification in which
each p-brane is wrapped on a p-cycle to end with a solution representing
an extreme black hole of the compactified supergravity theory.
In spite of the fact that it is useful for a wide range of intersecting 
brane configurations, several of the most interesting cases that recently 
appeared in the literature in the context of supersymmetric gauge theories
are excluded, namely, the localized overlappings (or intersections)
of p-branes.

~

Two overlapping branes can be
obtained from intersecting branes by separating each brane in a direction 
transverse to the remaining one. This sort of configurations are
included in the analysis of Ref.\cite{AEH}. However, as pointed out in 
Refs.\cite{PT,GKT,Ts3,Ga1}, this kind of solutions are not true
overlappings or intersections 
in the sense that the harmonic function corresponding
to a given brane is translational invariant in the directions
tangent to the other brane or, what is the same, each brane is
delocalized in its relative transverse space. 
Instead, we would like to consider in this paper p-brane
configurations with localized intersections which are, as we
briefly argued above, relevant for the study of strong coupling
phenomena in supersymmetric gauge theories. To our knowledge, the
first example of this kind of configuration was first constructed by 
Khuri \cite{K} while studying a string-like soliton solution 
in heterotic string theory, though the interpretation as a system of 
localized intersecting branes was not discussed there. 
Recently, Gauntlett, Kastor and Traschen \cite{GKT} have clarified this 
issue showing that it corresponds to two NS5-branes intersecting on a 
string in type II string theory. 
As also shown in Ref.\cite{GKT}, when uplifting
this solution to eleven dimensions, one is faced with a 
configuration of two M5-branes overlapping on a string that has
a striking characteristic: although the harmonic functions do
depend on the relative transverse directions, they are
translational invariant in the remaining overall transverse
direction. There may be solutions in which the
M5-branes are also localized in the overall transverse direction
but, if so, they shall not be given by the harmonic superposition
rule \cite{To3}. 

~

It is clear that, starting from this eleven dimensional 
configuration, a large class of solutions would be accesible,
whose distinguishing characteristic will be that all branes, while being 
localized in the relative transverse space, are delocalized in the 
overall transverse directions. This defines an appropriate kind of 
configurations in the context of the recently developed brane techniques 
for the study of non-perturbative phenomena and dualities of 
supersymmetric gauge theories \cite{HW,EGK,BSTY,BHOY,EJS,Witt}.
In fact, these techniques involve arrangements of flat p-branes with
{\em localized} intersections, such that the field theory on the
world-volume of the branes has the desired gauge symmetry, matter
content and degree of supersymmetry in the spacetime dimension which is
specifically chosen (three in \cite{HW}, four in the rest of the papers
cited above). In these approaches, one of
the constituent brane is finite in a given direction in which it is
stretched between other much heavier branes. To obtain explicit 
solutions of supergravity displaying this behaviour is, of course,
an involved task. However, a first approach in this direction is
given by finding generic configurations of infinite p-branes with 
localized intersections, that preserve a certain number of
supersymmetries. Indeed, as discussed in Ref.\cite{Ga1}, under
certain circumstances it is possible to think of these
configurations as corresponding to a p-brane stretched between
other branes.

~

In this paper we establish some rules for this kind of 
intersections. We find these rules purely from the bosonic
equation of motion of the low-energy theory, in the spirit
of Ref.\cite{AEH}. The rule corresponding
to localized intersections of p-branes is obtained in Section 2
by solving the equations of motion using an ansatz that
accounts for the properties of extremality and zero binding
energy. 
The whole set of classical fields is written in terms of a number of  
functions equal to the number of single branes involved in the
configuration. It also implies the preservation of a specific
amount of supersymmetry. 
In Section 3, we analize the resulting solutions
in ten and eleven dimensional theories and comment on
their relation with certain brane configurations which are relevant for
the study of non-perturbative supersymmetric
gauge theories. It is important to point out that {\em there are no 
new solutions within our ansatz}. 
Our rule manages to collect many of the solutions that 
have appeared in the literature\footnote{We should mention that
there are also solutions representing localized intersections of
p-branes that could not be reached from our starting ansatz 
({\it e.g.} those obtained in Ref.\cite{GGPT} from hyper--K\"ahler 
manifolds).} in a unique algebraic expression which is fully derived
within the bosonic sector of the theory. These solutions are connected
among themselves by a chain of dualities and dimensional reductions,
a fact which is at the root of the possibility to build a unique
expression that accounts for an entire family of solutions.
We think that most of the interest of our approach is
given by the fact that it provides a systematic alternative procedure
--with respect to the usual one relying on $\Gamma$-matrices algebra 
(see, for example, \cite{To3} and references therein)-- to build and
classify intersecting brane configurations. Indeed, generalizations of our
ansatz should lead to the appearance of new solutions as, {\em e.g.},
non-extremal branes, branes at angles, (p,q) webs, etc.

~

In Section 4, we show that a third brane can be added into
the configuration with vanishing binding energy. We derive the
corresponding intersection rule. We impose some new conditions on the
metric that reproduce the extremality nature of the configuration.
In Section 5, we analyze the solutions that emerge from these 
intersection rules, which again fit several known cases in the 
literature, and comment on their close relation with some
of the configurations used in the brane approach to
strong coupling phenomena of supersymmetric gauge theories. 
These configurations, as well as other pairwise intersections 
that can be obtained from them, are given in terms of a
non-harmonic function. We show that they provide a generalization
of the class of solutions discussed in Ref.\cite{AEH}, thus obeying 
the same intersection rule. 
Finally, in Section 6, we discuss our results and make some futher 
comments. In the Appendices, the detailed
expression for the Ricci tensor is given both for the two brane
and three brane cases, and the way the extremality condition 
appears in our framework is clarified.

\section{Localized Intersection of Two Branes}

Consider the following general expression for the bosonic sector of the
low-energy effective action corresponding to superstring theory in any
spacetime dimension $D$, $D \leq 11$,
\beq
S = \frac{1}{16\pi G_D} \int d^D x \sqrt{-g} 
\left( R - \frac{1}{2} (\partial \phi)^2 -
\sum_{A=1}^Q \frac{1}{2 n_A !}e^{a_A \phi}F_{n_A}^2 \right) ~, 
\label{action}
\eeq
The action includes gravity, a dilaton and $Q$ field strengths of 
arbitrary form degree $n_A \le D/2$ and coupling to the dilaton $a_A$.
The metric is expressed in the Einstein frame. 
There may be Chern-Simons terms in the action but we omit them as they are
irrelevant for the kind of solutions we will concentrate on.
Although we take the spacetime 
to have a generic dimension $D$, this action is most suitable for describing 
the bosonic part of $D=10$ or $D=11$ supergravities. 
The equations of motion can be written in the following form:
\beq
{R^\mu}_\nu = \half\partial^\mu \phi \partial_\nu \phi + \sum_{A=1}^Q
{{\Theta_A}^{\mu}}_{\nu} ~,
\label{einstein}
\eeq
\beq
\frac{1}{\sqrt{-g}}\dmu(\sqrt{-g}\dmup)\phi =\sum_{A=1}^Q 
\frac{a_A}{2n_A!}e^{a_A\phi}F_{n_A}^2 ~,
\label{dilaton}
\eeq
\beq
\partial_{\mu_1}\left(\sqrt{-g}e^{a_A\phi}F^{\mu_1\ldots\mu_{n_A}}
\right)=0 ~,
\label{maxwell}
\eeq
where ${{\Theta_A}^{\mu}}_{\nu}$ is the contribution to the 
stress-energy tensor corresponding to the $n_A$-form,
\beq
{{\Theta_A}^{\mu}}_{\nu} =
\frac{1}{2n_A!}e^{a_A\phi} \left(n_A F^{\mu\rho_2\ldots\rho_{n_A}}
F_{\nu\rho_2\ldots\rho_{n_A}}- \frac{n_A-1}{D-2} F_{n_A}^2 
{\delta^\mu}_\nu \right) ~. 
\label{stress}
\eeq
We must supplement the equations of motion by imposing the Bianchi 
identities to the $n_A$-forms,
\beq
\partial_{[\mu_1}F_{\mu_2\ldots\mu_{n_A+1}]}=0 ~. 
\label{bianchi}
\eeq
as they are field strengths of $(n_A-1)$-form potentials. We are 
interested in classical solutions describing a pair of p-branes that are
translationally invariant in the overall transverse directions but are
localized in the relative transverse coordinates. Thus, we set (for
simplicity) all but two field strengths to zero (a condition that will
be relaxed in Section 4).

~

Let us now specialize to a particular form of the metric which is a
slight generalization of the p-brane ansatz, and lead us to obtain the 
class of configurations we want to deal with. 
The line element is given by 
\bea
ds^2 & = & -\,B^2 dt^2 + C^2 \delta_{ij} ds^i ds^j + X^2 \delta_{ab} 
dx^a dx^b + Y^2 \delta_{\a\b} dy^\a dy^\b \nonumber \\
& & +\,W^2 \delta_{\mu\nu} dw^{\mu} dw^{\nu} ~,
\label{metric1}
\eea
where the $s_i$'s span the intersection, $i,j=1\ldots \bar{q}$, the $x_a$'s 
and $y_\a$'s are the relative transverse coordinates $a,b=1\ldots p_1$ and 
$\a,\b=1\ldots p_2$, whereas the $w_{\mu}$'s are the overall transverse 
coordinates $\mu = 1\ldots p_t$. 
The functions $B$, $C$, $X$, $Y$ and $W$ depend only on the 
relative transverse coordinates $x^a$, $y^\a$. 
It is clear that the relation $\bar{q}+p_1+p_2+p_t=D-1$ must be satisfied. 
Furthermore, we will consider solutions that allow a factorization of the 
form:
\beq
{\cal F}(x^a,y^\a) \equiv {\cal F}_x(x^a){\cal F}_y(y^\a) ~,
\label{factor}
\eeq
for the whole set of functions. These solutions will represent a $q_1$-brane
and a $q_2$-brane with a $\bar{q}$-dimensional localized intersection, being 
$q_A = p_A +\bar{q}$.

~

For the $n_A$-form field strengths, we can generally make two
kinds of ans\"{a}tze. The electric ansatz is done asking that the Bianchi
Identities are trivially satisfied. Consider, for example, an electrically 
charged $q_1$-brane. It couples naturally to a $q_1+2$-form, $F_{n_1}$,
\beq
F_{0 i_1 \ldots i_{\bar{q}} a_1 \ldots a_{p_1} \a} =
\epsilon_{i_1 \ldots i_{\bar{q}}} \epsilon_{a_1 \ldots a_{p_1}}
\derya E_1 ~.
\label{electric}
\eeq
For a magnetic brane, on the other hand, one asks that the equations of 
motion for the field strength (\ref{maxwell}) are trivially satisfied. 
Thus, a magnetically charged $q_1$-brane couples to a $(D-q_1-2)$-form that
can be written as follows
form,
\beq
F^{\mu_1 \ldots \mu_{p_t} \a_1 \ldots \a_{p_2-1}} = \frac{1}{\sqrt{-g}}
e^{-a\phi}\epsilon^{\mu_1 \ldots \mu_{p_t}} \epsilon^{\a_1 \ldots \a_{p_2-1} \a}
\derya E_1 ~
\label{magnetic}
\eeq
We remark here on the fact that the derivative is always taken with respect 
to directions which are {\it perpendicular} to the respective brane. This is
in accordance with the cases considered in \cite{GKT} and with the
ans\"{a}tze of \cite{AEH} for intersecting branes. 
Also the dilaton depends on the relative transverse coordinates.

~

Let us now discuss in some detail the next ans\"{a}tze that we will make
in order to solve the equations of motion (\ref{einstein})--(\ref{maxwell}).
The Ricci tensor that corresponds to the metric (\ref{metric1}) (which is
computed in an Appendix) displays several terms that mix non-trivially the
different components of the metric. On the other hand, we should be able to 
write $B$, $C$, $X$ and $Y$ in terms of a pair of functions
in order to make the distinction between the constituents
branes. These functions are supplemented by the metric component relative
to the overall transverse space $W$, the dilaton $\phi$ and the set of
functions $E_A$. We are thus forced to impose further constraints on our
configuration. If viewed as a classical configuration of supergravity, the
conditions that one should impose would be the vanishing of the 
supersymmetry transformations --corresponding to a given infinitesimal 
parameter $\eta$ that satisfies certain chirality constraints-- for all 
the fermions. This amounts to the preservation
of some of the supersymmetries (those related to the particular parameter
$\eta$). In our approach, we are not going to analyze the whole content
of the supergravity theory that is behind (\ref{action}). We should,
instead, take profit of the signals left into the purely bosonic 
configuration by the existence of certain unbroken supersymmetries, 
that is, the vanishing binding energy for the overlapping brane
configuration. To this end, we impose that the mixing terms of the
Ricci tensor mentioned above vanish \cite{AEH}. This happens provided 
that the following constraints are imposed:
\beq
B_xC_x^{\bar{q}}X_x^{p_1-2}Y_x^{p_2}W_x^{p_t} = 1 ~,
\label{const1}
\eeq
\beq
B_yC_y^{\bar{q}}X_y^{p_1}Y_y^{p_2-2}W_y^{p_t} = 1 ~.
\label{const2}
\eeq
These constraints can be physically interpreted as enforcing extremality.
They can be read as a way to express $W$ as a function of the other metric
components. It is also interesting to point out now, that all the
solutions investigated in Refs.\cite{GKT,Ts3,Ga1} satisfy them, as will be
better described below.

~

\noindent
After (\ref{const1}) and (\ref{const2}), we can rewrite Einstein equations as
\beq
\Box\ln{B} = \frac{1}{2\sqrt{-g}} 
\sum_{A=1}^2 \frac{\xi_A^0}{D-2} S_A (\nabla_{\top} E_A)^2 ~,
\label{ein00}
\eeq
\beq
\Box\ln{C} = \frac{1}{2\sqrt{-g}}
\sum_{A=1}^2 \frac{\xi_A^s}{D-2} S_A (\nabla_{\top} E_A)^2 ~,
\label{einij}
\eeq
\bea
& \delta_{ab} \Box\ln{X} + X^{-2} \left[(p_1 
- 2)\derxa\ln{X}\derxb\ln{X} + p_2\derxa\ln{Y}\derxb\ln{Y} \right. \nonumber \\
& + \left. \bar{q}\derxa\ln{C}\derxb\ln{C} + p_t\derxa\ln{W}\derxb\ln{W} + 
\derxa\ln{B}\derxb\ln{B} \right] = \nonumber \\
& = \frac{1}{2\sqrt{-g}} \left[ \sum_{A=1}^{2} \frac{\xi_A^x}{D-2} 
S_A (\nabla_{\top} E_A)^2 \delta_{ab} + S_2 (\derxa E_2)(\derxb E_2) \right] ~,
\label{einab}
\eea
\bea
& \delta_{\a\b} \Box\ln{Y} + Y^{-2} \left[p_1\derya\ln{X}\deryb\ln{X} 
+ (p_2 - 2)\derya\ln{Y}\deryb\ln{Y} \right. 
\nonumber \\
& + \left. \bar{q}\derya\ln{C}\deryb\ln{C} + p_t\derya\ln{W}\deryb\ln{W} + 
\derya\ln{B}\deryb\ln{B} \right] = \nonumber \\
& = \frac{1}{2\sqrt{-g}} \left[ \sum_{A=1}^2 \frac{\xi_A^y}{D-2} S_A 
(\nabla_{\top} E_A)^2 \delta_{\a\b} + S_1 (\derya E_1)(\deryb E_1) \right] ~, 
\label{einpapb}
\eea
\bea
& (p_1-2)\derxa\ln{X}\deryb\ln{X} + (p_2-2)\derxa\ln{Y}\deryb\ln{Y}
+ 2 \derxa\ln{Y}\deryb\ln{X} \nonumber \\
& + \bar{q}\derxa{C}\deryb{C} + \derxa{B}\deryb{B} + {p_t}\derxa{W}\deryb{W}
= 0 ~,
\label{einapb}
\eea
\beq
\Box\ln{W} = \frac{1}{2\sqrt{-g}}
\sum_{A=1}^2 \frac{\xi_A^w}{D-2} S_A (\nabla_{\top} E_A)^2 ~,
\label{einmunu}
\eeq
where we have introduced the symbol $\nabla_{\top}E_A$ to refer to the gradient
of $E_A$ with respect to coordinates relatively transverse to the $q_A$-brane. 
It is worth noting that this does not mean at all that functions $E_A$ depend 
only on those coordinates, as will be clear below. We have also introduced
the D'Alembertian, which after (\ref{const1}) and (\ref{const2}) is simply,
\beq
\Box = Y^{-2}\dery^2 + X^{-2}\derx^2.
\label{Dal}
\eeq
and the quantities $\xi_A^z$ ($z$ being the label for the different blocks of
coordinates), whose value is given by
\beq
\xi_A^z = \left\{ \begin{array}{l} D-q_A-3 \;\;\; \mbox{if $z$ is 
longitudinal to the $q_A$-brane} ~, \\ -(q_A+1) \;\;\; \mbox{if $z$ 
is transverse to the $q_A$-brane} ~. \end{array} \right.
\label{defin1}
\eeq
Finally, we have defined for convenience the quantities $S_A$,
\beq
S_1 = \frac{W^{2p_t}Y^{2(p_2-1)}}{\sqrt{-g}}e^{\e_1a_1\phi} ~,
\label{s1}
\eeq
\beq
S_2 = \frac{W^{2p_t}X^{2(p_1-1)}}{\sqrt{-g}}e^{\e_2a_2\phi} ~,
\label{s2}
\eeq
where $\e_A$ is a positive sign for the electric membranes and a negative sign for
the magnetic ones. Note that the metric determinant has a very simple form as 
a consequence of the `no-force' conditions (\ref{const1})--(\ref{const2})
imposed to the metric,
\beq
\sqrt{-g} = X_x^2Y_y^2 ~.
\label{det}
\eeq
We must still impose the equations of motion corresponding to the dilaton,
\beq
\Box\phi = -\frac{1}{2\sqrt{-g}}
\sum_{A=1}^2 \e_A a_A S_A (\nabla_{\top} E_A)^2 ~,
\label{dilat}
\eeq
and the $q_A$-forms,
\beq
\derya ( S_1 \derya E_1) = \derxa ( S_1 \derya E_1) = 0 ~,
\label{q1form}
\eeq
\beq
\derxa ( S_2 \derxa E_2) = \derya ( S_2 \derxa E_2) = 0 ~,
\label{q2form}
\eeq
We will finally consider the following ansatz\footnote{We present 
eqs.(\ref{e1})--(\ref{s2x}) as an ansatz for simplicity. It is possible 
to argue that they are forced by the equations of motion
following the lines of Ref.\cite{Oh}.}, 
\beq
E_1 = l_1{S_1}_x^{-1}H_1^{-1} ~,
\label{e1}
\eeq
\beq
E_2 = l_2{S_2}_y^{-1}H_2^{-1} ~,
\label{e2}
\eeq
\beq
{S_1}_y = H_1^2 ~,
\label{s1y}
\eeq
\beq
{S_2}_x = H_2^2
\label{s2x}
\eeq
(with $l_1$ and $l_2$ a couple of --up to now-- arbitrary constants),
that enables us to write everything in terms of a pair of functions $H_A$
which, after eqs.(\ref{q1form})--(\ref{q2form}), must be
harmonic,
\beq
\dery^2 H_1 = \derx^2 H_2 = 0 ~.
\label{harmonic}
\eeq
In this way, our solutions are going to be characterized by a pair of
harmonic functions corresponding to the same number of constituents
branes that participate on the configuration. From the point of
view of supergravity,
this shall mean that we are dealing with configurations that preserve
one quarter of the supersymmetries. The most general solutions to
eqs.(\ref{harmonic}) are given by an arbitrary set of superpositions
of identical branes {\em localized} along the relative transverse directions:
\beq
H_1 = 1 + \sum_j \frac{c_j}{|\vec{y}-\vec{y}_j|^{p_2-2}} ~,
\label{harm1}
\eeq
\beq
H_2 = 1 + \sum_j \frac{d_j}{|\vec{x}-\vec{x}_j|^{p_1-2}} ~.
\label{harm2}
\eeq
This is the well-known multicenter solution whose existence is due to
the no-force condition \cite{Ts2} that we previously imposed in
(\ref{const1})--({\ref{const2}). Now, if we demand the set of conditions,
\beq
W_x^{2p_t}Y_x^{2(p_2-2)}e^{\e_1a_1\phi_x} = 1 ~,
\label{last1}
\eeq
\beq
W_y^{2p_t}X_y^{2(p_1-2)}e^{\e_2a_2\phi_y} = 1 ~,
\label{last2}
\eeq
it is quite easy to solve the dilaton equation of motion,
\beq
\phi = \sum_{A=1}^2 \e_A a_A \alpha_A \ln H_A ~,
\label{dil}
\eeq
as well as the diagonal components of the Einstein equations,
\beq 
\ln B = \ln C = - \sum_{A=1}^2 \frac{D-q_A-3}{D-2} \alpha_A \ln H_A ~, 
\label{ein1}
\eeq
\beq
\ln X = - \sum_{A=1}^2 \frac{\xi^x_A}{D-2} \alpha_A \ln H_A ~, 
\label{ein2}
\eeq
\beq
\ln Y = - \sum_{A=1}^2 \frac{\xi^y_A}{D-2} \alpha_A \ln H_A ~, 
\label{ein3}
\eeq
\beq 
\ln W = \sum_{A=1}^2 \frac{q_A+1}{D-2} \alpha_A \ln H_A ~, 
\label{ein4}
\eeq
provided $\a_A = \frac{1}{2} l_A^2$. The fact that $B$ and $C$ are equal 
could have
been directly predicted from the SO$(1,\bar{q})$ boost invariance of the
extremal configuration. The only equations that remain to be solved are
the off-diagonal Einstein equations which at this stage simply reduce to
a set of algebraic equations:
\beq
\left[ \frac{(p_2 + p_t - 2)(q_1 + 1)^2 + (q_1 + 1)(D - q_1 - 3)^2}{(D
- 2)^2} + \half a_1^2 \right] \alpha_1 = 1 ~,
\label{algeb1}
\eeq
\beq
\left[ \frac{(p_1 + p_t - 2)(q_2 + 1)^2 + (q_2 + 1)(D - q_2 - 3)^2}{(D
- 2)^2} + \half a_2^2 \right] \alpha_2 = 1 ~,
\label{algeb2}
\eeq
and
\bea
\sum_{A \neq B = 1}^2 (\bar{q} + 3)(D - q_A - 3)(D - q_B - 3) + 
p_t(q_A + 1)(q_B + 1) \nonumber \\
- 2(p_A - 2)(D - q_A -3)(q_B + 1) + \half (D - 2)^2 \e_A a_A \e_B 
a_B = 0 ~.
\label{algeb3}
\eea
From the first two equations, we obtained an explicit value for
the $\a_A$'s in terms of the dimensions of the constituent branes 
and the dilaton couplings
\beq
\a_A = \frac{D-2}{\Delta_A} ~,
\label{alphas}
\eeq
where
\beq 
\Delta_A = (q_A + 1)(D - q_A - 3) + \half (D - 2) a_A^2 ~, 
\label{Delta}
\eeq
that coincides with the one obtained in the case studied in 
Ref.\cite{AEH}. The third equation leads us directly to the announced 
{\it rule for localized intersections of p-branes} which is one of the
main results of this paper:
\beq
\bar{q} + 3 = \frac{(q_1+1)(q_2+1)}{D-2} - \half \e_1\e_2 a_1a_2 ~.
\label{overrule}
\eeq
This expression is very similar to that of Ref.\cite{AEH}, except for 
a shift of two units in $\bar{q}$. 
We will show in the next Section, that this rule 
leads to most known cases of localized intersections of p-branes
appearing in the literature (within the scope of our ansatz), and 
that it does not have further solutions. In that respect, we think
that the main interest of eq.(\ref{overrule}) relies in the fact that
it is a unique algebraic expression that collects a complete family
of solutions which are related by various dualities and dimensional
reductions.

~
 
Let us close this Section by stressing that the configuration we have
obtained so far, 
\beq 
B = C = \prod_{A=1}^2 H_A^{-(D - q_A - 3)/\Delta_A} ~,
\label{solut1}
\eeq
\beq
X = \prod_{A=1}^2 H_A^{-\xi^x_A/\Delta_A} ~, 
\label{solut2}
\eeq
\beq
Y = \prod_{A=1}^2 H_A^{-\xi^y_A/\Delta_A} ~, 
\label{solut3}
\eeq
\beq 
W = \prod_{A=1}^2 H_A^{(q_A + 1)/\Delta_A} ~,
\label{solut4}
\eeq
\beq
e^\phi = \prod_{A=1}^2 H_A^{\e_A a_A (D - 2)/\Delta_A} ~,
\label{soldil}
\eeq
\beq
E_1 = \sqrt{2\alpha_1} \, H_{1}^{-1}
H_{2}^{-2\frac{\xi_{2}^{x} - \xi_{2}^{y}}{\Delta_{2}}} ~,
\eeq
\beq
E_2 = \sqrt{2\alpha_2} \, H_{1}^{2\frac{\xi_{1}^{x} - 
\xi_{1}^{y}}{\Delta_{1}}} H_{2}^{-1} ~,
\eeq
is consistent with the conditions (\ref{last1}) and (\ref{last2}), and
obeys the harmonic superposition rule \cite{Ts}.

\section{Localized intersections in various dimensions}

In this section, we study the complete set of solutions admitted
by eq.(\ref{overrule}) in various
spacetime dimensions. We stress on the duality chains that
relate different solutions among themselves. We will use a
very convenient notation introduced in Ref.\cite{To3}, denoting
by $(\bar{q}|{\cal M}_1,{\cal M}_2)$ to the localized overlapping
or intersection between an
${\cal M}_1$-brane and an ${\cal M}_2$-brane with $\bar{q}$
common tangent directions. We will see that the solutions are not 
new but they correspond to a family of configurations that have
been separately considered before by many authors.

\subsection{Localized intersections of M-branes}

Let us first analyze the eleven dimensional case. There is a 4-form
field strength in $D=11$ supergravity, that can describe either 
electric M2-branes or magnetic M5-branes. As there is no dilaton, we
simply put $a_A=0$ for the 4-forms. Then, the overlapping rule 
derived in eq.(\ref{overrule}) aquires the simpler form:
\beq
\bar{q} + 3 = \frac{(q_1+1)(q_2+1)}{9} ~,
\label{rule11d}
\eeq
whose only solution is $q_1 = q_2 = 5$ and $\bar{q} = 1$, that is, 
M5-branes overlapping in a string, $(1|M5,M5)$. 
This is exactly the solution
recently obtained in Ref.\cite{GKT} by rather different means. 
There, the solution is found by uplifting either overlapping
NS5-branes or mutually orthogonal D4-branes to eleven dimensions. 

Explicitely, we find $\a_A = 1/2$, and a line element given by
\beq
ds^2 = H_1^{2/3}H_2^{2/3}\left(H_1^{-1}H_2^{-1}(-dt^2 + ds^2) +
H_1^{-1}d\vec{x}^2 + H_2^{-1}d\vec{y}^2 + d\omega^2\right) ~.
\label{1|m5m5}
\eeq
As explained in Ref.\cite{Ga1}, this solution does not satisfy
the (p-2)-dimensional self-intersection rule for p-branes
\cite{PT}. This puzzle is solved by observing that a third
brane can be introduced without breaking further 
supersymmetries \cite{Ts3,Ga1}. We will discuss this point
in the next Section, where we will obtain a general rule
for the introduction of a third brane inside an overlapping
configuration. It would also be interesting to explore what
kind of solution appears if localization in the 
$\omega$-direction is demanded.

\subsection{Localized intersections of NS-branes with other branes}

The NS-NS sector of string theory is known to posses a 3-form
field strength that couples to the dilaton with $a_A=-1$. It can
couple to a fundamental string and to a magnetic NS5-brane.
The overlapping rule for these objects is:
\beq
\bar{q} + 3 = \frac{(q_1+1)(q_2+1)}{8} - \half \e_1\e_2 ~.
\label{rule10dNSNS}
\eeq
It is immediate to see that this equation admits only one solution 
describing two NS5-branes overlapping in a string, $(1|NS5,NS5)$. 
This solution can also be obtained from a string-like soliton solution 
of heterotic string theory first considered in Ref.\cite{K}, 
by setting the gauge fields to zero \cite{GKT}. Once again 
$\a_A = 1/2$ and the line element is simply:
\beq
ds^2 = H_1^{3/4}H_2^{3/4}\left(H_1^{-1}H_2^{-1}(-dt^2 + ds^2) +  
H_1^{-1}d\vec{x}^2 + H_2^{-1}d\vec{y}^2\right) ~.
\label{1|ns5ns5}
\eeq
A common feature of these solutions is the appearance of
an overall conformal factor, while each direction tangent to the
worldvolume of a $q_A$-brane gets a factor $H_A^{-1}$ as already
noticed in Refs.\cite{Ts3,Ga1,To3}.

~

Now we look at the localized intersections of NS-branes and D-branes.
We use the fact that $a_A=(5-n_A)/2$ is the coupling to the 
dilaton of the $n_A$-form field strengths coming from the 
RR sector. It is immediate to see that equation (\ref{overrule})
does not admit localized intersections of a fundamental string and 
the D-branes. Concerning NS5-branes, the overlapping rule for
these objects and Dq-branes is:
\beq
\bar{q} = q - 3 ~,
\label{rule10dNSD}
\eeq
which precisely agrees with the case by case result obtained in
Ref.\cite{Ga1}. So, the possible overlappings between NS5-branes 
and D-branes are $(q-3|NS5,Dq)$, for $3 \le q \le 8$.

~

Here we can give two important examples of such overlapping which
are going to be better clarified in Section 5.
The first one is that appearing in the brane setup of Ref.\cite{HW} 
for a configuration which gives $N = 2$ supersymmetry in 
3 spacetime dimensions. It consists of a NS5-brane with (12345) spatial
directions and a D5-brane with (12789) spatial directions such that
the number of common directions are coincident with our previously
derived rule (\ref{rule10dNSD}). One would, in principle, have
expected that this configuration matches the overlapping rule
because, as explained in Ref.\cite{HW}, it is crucial there that
both membranes are well-localized in their relative transverse
directions. They are also localized in the overall transverse
$x_6$-direction.

~

The second example appears in the brane setup of Ref.\cite{EGK} for a
configuration which gives $N = 1$ supersymmetry in 4 spacetime
dimensions. In this configuration there are two types of NS5-branes,
one denoted by NS5 in the (12345) spatial directions and one denoted 
by NS5' in the (12389) spatial directions. There are D6-branes in the 
(123789) spatial directions. The rule (\ref{overrule}) accounts for the
localized intersection between the D6-branes and the NS5-branes, but seems
to disagree with the intersection between the D6-branes and the NS5'-brane.
Moreover, this last intersection obeys the rule (33) of Ref.\cite{AEH}.  
In the next Section, we will show that --for the case of localized 
intersections-- we can add new branes with vanishing binding energy whose 
intersection rule is not given by (\ref{overrule}). The new intersection 
rule is precisely that of equation (33) in Ref.\cite{AEH}, and the origin
of this coincidence will be clarified.

\subsection{Localized Intersections of D-branes}

The D-branes are the bearers of the RR charges and they give rise to
field strengths which couple to the dilaton in such a way that 
$\epsilon_A a_A = (3-q_A)/2$ both for electrically and magnetically
charged $q_A$-branes. Then the equation
(\ref{overrule}) gives:
\begin{equation}
2 \bar{q} + 8 = q_{1} + q_{2} ~.
\label{rule10dDD}
\end{equation}
The D-branes shall intersect in such a way that there are eight relative
transverse directions, a condition which is also known to be required by
the preservation of unbroken supersymmetries \cite{PCJ,BBJ}.
The solution $(0|D4,D4)$ has the following line element,
\beq
ds^2 = H_1^{5/8}H_2^{5/8}\left(H_1^{-1}H_2^{-1}(-dt^2) +  
H_1^{-1}d\vec{x}^2 + H_2^{-2}d\vec{y}^2 + d\omega^2\right) ~.
\label{0|d4d4}
\eeq
It could have also been obtained just by dimensional reduction of
$(1|M5,M5)$ on the common string direction. One can obtain the
remaining type II solutions by a chain of T dualities,
\beq
\begin{array}{ccc}
(0|D4,D4) & \tdual & (0|D3,D5) \\
\tdua & & \tdua \\
(1|D5,D5) & \tdual & (1|D4,D6) \\
\end{array} ~.
\label{dbr1}
\eeq
There are more configurations that are,
in principle, accesible by this procedure,
\beq
\begin{array}{ccccc}
(0|D2,D6) & \tdual & (0|D1,D7) & \tdual & (0|D0,D8) \\
\tdua & & \tdua & & \tdua \\
(1|D3,D7) & \tdual & (1|D2,D8) & \tdual & (1|D1,D9)
\end{array}
\label{dbr2}
\eeq
We present them separately for the following reason: the relative transverse 
space of one of the D-branes, being less or equal than two-dimensional, 
leads to the appearance of logarithmic or linear singularities.
Consider, for example, $(1|D3,D7)$ and see that the harmonic function
corresponding to the D7-brane is harmonic in the Euclidean plane. So,
each D7-brane produces a conical singularity and its energy per unit
7-volume result to be logarithmically divergent. It is also interesting
to mention the case $(1|D2,D8)$, where an harmonic function in one
Euclidean coordinate corresponds to the D8-branes thus being piecewise
linear\footnote{The D8-brane is a domain wall solution of a massive 
version of type IIA supergravity, separating regions of different 
cosmological constants \cite{PW,BRGPT}}. Following a similar reasoning, 
it is immediate to see that the configuration $(1|D1,D9)$ is equivalent
to an isolated D1-brane in empty ten dimensional Minkowski space.
In the context of supergravity, all these solutions are known to preserve 
1/4 of the original supersymmetries \cite{Ts3,Ga1}. 
They are related to the configurations obtained in the previous 
subsection by S-duality.
 
\section{Adding a third localized p-brane}

Let us consider the possibility of adding a third brane into the
picture with the following two requirements:
 
(i) the brane spans the intersecting and totally transverse 
coordinates, thus being a $(\bar{q}+p_t)$-brane.
 
(ii) the geometry of the target space gets modified only by
the introduction of a third function $H_3$, that corresponds
to the new brane, while the contributions of the overlapping
branes are untouched. This is analogous to the `no-force'
condition, in the sense that we shall add a new object that 
modifies the geometry without excerting
any gravitational attraction to the previous configuration.
This requirement is related to the existence of a certain
amount of unbroken supersymmetries or, in our case, to the
elimination of mixing terms in the Ricci tensor.

~

\noindent
We define new functions for the line element 
\bea
ds^2 & = & -\,\hat{B}^2 dt^2 + \hat{C}^2 \delta_{ij} ds^i ds^j + 
\hat{X}^2 \delta_{ab} dx^a dx^b + \hat{Y}^2 \delta_{\a\b} dy^\a dy^\b \nonumber \\
& & +\,\hat{W}^2 \delta_{\mu\nu} dw^{\mu} dw^{\nu} ~,
\label{metric2}
\eea
such that the contribution of the new brane 
is given by a new factor $\tilde{\cal F}(x^a,y^\a)$,
\beq
\hat{\cal F}(x^a,y^\a) \equiv {\cal F}(x^a,y^\a) 
\tilde{\cal F}(x^a,y^\a) = {\cal F}_x(x^a){\cal F}_y(y^\a)
\tilde{\cal F}(x^a,y^\a) ~,
\label{factor2}
\eeq
that we allow to depend on the whole set of relatively
transverse coordinates. Note the difference with respect to the previous 
case where we imposed a factorization (\ref{factor}) of the
coordinates dependence. Here, the coordinates which are relatively 
transverse to the third brane are $x_a$'s and $y_\a$'s.

~

In order to see how our requirements (i) and (ii) severely
constrain the resulting configuration, we compute the Ricci tensor 
(see Appendix) in the background of the former intersecting brane
solution, and impose the `no-force' conditions --thought of
as the vanishing of the mixing terms in the Ricci tensor-- as before,
\beq
\tB\tC^{\bar{q}}\tX^{p_1-2}\tY^{p_2}\tW^{p_t} = 1 ~,
\label{const3}
\eeq
\beq
\tB\tC^{\bar{q}}\tX^{p_1}\tY^{p_2-2}\tW^{p_t} = 1 ~.
\label{const4}
\eeq
From these conditions, it is immediate to see that
\beq
\tX^2 = \tY^2
\label{const5}
\eeq
and
\beq
\tB\tC^{\bar{q}}\tX^{p_1+p_2-2}\tW^{p_t} = 1 ~.
\label{const6}
\eeq
By plugging these conditions back into the Ricci tensor, we can write the
equations of motion, in the background of the former overlapping 
brane configuration. The introduction of the third brane modifies
the $S_A$ functions, $S_A \to \hat{S}_A$ with
\beq
\hat{S}_1 = \tW^{2p_t}\tX^{2(p_2-2)} e^{\e_1 a_1 \tilde{\phi}} S_1 ~,
\label{hatese1}
\eeq
\beq
\hat{S}_2 = \tW^{2p_t}\tX^{2(p_1-2)} e^{\e_2 a_2 \tilde{\phi}} S_2 ~,
\label{hatese2}
\eeq
where $\tilde{\phi}$ is the correction to the dilaton also due to 
the third brane. The only way to solve our new system without 
changing the values of $H_{1}$ and $H_{2}$, or the ones of $S_{1}$ and 
$S_{2}$, is to impose the following conditions:
\beq
\tW^{2p_t}\tX^{2(p_2-2)} e^{\e_1 a_1 \tilde{\phi}} = 1 ~,
\label{more1}
\eeq
\beq
\tW^{2p_t}\tX^{2(p_1-2)} e^{\e_2 a_2 \tilde{\phi}} = 1 ~,
\label{more2}
\eeq
which also lead to an expression of the dilaton in terms of the
metric function $\tX$,
\beq
\tX^{2(p_{1}-p_{2})} = e^{(\e_1 a_1 - \e_2 a_2)\tilde{\phi}} ~.
\label{more3}
\eeq
Now, the diagonal components of the Einstein equations for a
generic function ${\cal F} = B, C, W, X, Y$ are simply given by
\beq
\tBox\ln\tilde{\cal F} = 
\frac{1}{2\sqrt{-\hat{g}}} \frac{\xi_{\cal F}}{D-2} \hat{S}_3 
\left[ \hat{X}^{-2}(\derx E_3)^2 + \hat{Y}^{-2}(\dery E_3)^2 \right] ~,
\label{diagonal}
\eeq
where,
\beq
\hat{S}_{3} = \frac{\hat{X}^{2p_{1}}\hat{Y}^{2p_{2}}}{\sqrt{-\hat{g}}} ~,
\label{ese3}
\eeq
while the contribution of the third brane to each metric component is given by
the factor $\xi_{\cal F}$,
\beq
\xi_{\cal F} = \left\{ \begin{array}{l} D-(\bar{q}+p_t)-3 \;\;\; \mbox{if ${\cal F}$ 
is $B$, $C$ or $W$} ~, \\ -(\bar{q}+p_t+1) \;\;\; \mbox{if ${\cal F}$ is $X$ or $Y$} ~, 
\end{array} \right.
\label{defin2}
\eeq
in accordance to the previously established recipe (\ref{defin1}).
In order to solve (\ref{diagonal}), we should
set $E_{3} = l_{3} H_{3}^{-1}$ and $\hat{S}_{3} = H_{3}^{2}$.
Now, the solution is
\beq
\ln\tilde{\cal F} = - \frac{\xi_{\cal F}}{D - 2} \ln H_{3}^{l_3^2/2} ~,
\label{diagsol}
\eeq
if and only if $H_3$ satisfies 
\beq
(H_{2}^{-1} \nabla_{x}^{2} + H_{1}^{-1} \nabla_{y}^{2}) H_{3} = 0 ~,
\label{h3cond}
\eeq
thus being non-harmonic\footnote{Note that eq.(\ref{h3cond}) is indeed a 
curved space Laplace equation. Thus, though $H_3$ is not harmonic in the flat 
space sense, it can be thought of to be harmonic in some generalized curved
space sense. In fact, this equation appears whenever the effective transverse 
space is curved \cite{Ts4}. We are grateful to Arkady Tseytlin for his 
clarifying comments on this issue.}. 
This equation coincides with the one obtained in 
Refs.\cite{Ts3,Ga1} for the case of three branes in the context of supergravity.
It was also obtained earlier in the context of ten dimensional solutions of 
string theory representing extreme dyonic black holes \cite{Ts4}.

~
 
As before, the remaining block-diagonal Einstein equations give the value of
$\a_3=l_3^2/2$ whose expression is coincident to that of $\a_A$'s previously
obtained 
\beq
\a_3 = \frac{D-2}{\Delta_3} ~,
\label{alpha3}
\eeq
where
\[ \Delta_3 = (q_3 + 1)(D - q_3 - 3) + \half (D - 2) a_3^2 ~. \]
Here, we introduced the dimension of the third brane $q_{3} = \bar{q} + 
p_{t}$. On the other hand, the off-block-diagonal Einstein equations lead 
to the {\em intersection rule for the third brane},
\beq
\bar{q} + 1 = \frac{(q_{i} + 1)(q_{3} + 1)}{D-2} - \frac{1}{2} a_{i} a_{3}
\epsilon_{i}\epsilon_{3} ~,
\label{dimint3}
\eeq
where $i = 1, 2$ refers to anyone of the `old' two branes. Again, we
should say that the unique algebraic expression to which we arrive is
a fingerprint of the various dualities and dimensional reductions that
relate the solutions of (\ref{dimint3}) among themselves. We must comment
on the fact that the expression for the intersecting dimension is the same as 
the one obtained in Ref.\cite{AEH} for intersecting branes that depend in the
overall transverse coordinates. The reason of this coincidence will be
clarified below.

\section{Examples With Three Branes}

In this Section, we will study the three brane configurations that
solve eqs.(\ref{overrule}) and (\ref{dimint3}). We will introduce a simple 
notation that generalize that of Ref.\cite{To3}: a configuration corresponding 
to a localized intersection of dimension $\bar{q}$ of an ${\cal M}_1$-brane and 
an ${\cal M}_2$-brane, forming a bound state of zero binding energy with a
third brane ${\cal M}_3$, will be denoted as $({\cal M}_3|\bar{q}|{\cal 
M}_1, {\cal M}_2)$. Note that, as discussed above, the fully localized function 
corresponding to the ${\cal M}_3$-brane is, in principle, non-harmonic.
We will see that no new solutions emerge from our approach.
We will consider certain examples, which are related to well-known
brane configurations used to obtain information about
dualities and strong coupling effects in supersymmetric gauge theories.

\subsection{Three M-branes}

We start with the configuration introduced in Refs.\cite{Ts3,Ga1} and 
reobtained in section 3.1, i.e. with two M5-branes, one in (12345) spatial
directions and the other one in (16789) directions. 
Here, the overall transverse direction is $x_{10}$ and the 
intersecting direction is $x_{1}$. Thus, the third brane should be an
M2-brane that spans (1~10) spatial directions.
For $q_{1} = q_{2} = 5, q_{3} = 2$ and $a_{1} = a_{2} = a_{3} = 0$
the condition (\ref{dimint3}) is automatically satisfied.
The function corresponding to the third brane depends now on both
sets of variables which describe the wordvolumes of the two M5-branes,
so we have $H_{3}$ as a function of $(x^{2},\cdots , x^{9})$ that
satisfies the equation (\ref{h3cond}) which is just the same as 
the condition (3.8) in \cite{Ts3} or (25) in \cite{Ga1}.
By using the notations of formula (\ref{1|m5m5}) for the groups of
coordinates, the line element that corresponds to the $(M2|1|M5,M5)$
configuration is given by:
\bea
ds^{2} & = & H^{2/3}_{1}H_{2}^{2/3}H_{3}^{1/3} \left[ H_1^{-1}H_2^{-1}H_3^{-1}
(-dt^2 + ds^{2}) + H_{1}^{-1} d{\vec{x}}^{2} + H_{2}^{-1} d{\vec{y}}^{2}
\right. \nonumber \\ & + & \left. H_{3}^{-1} d w^{2} \right] ~.
\label{m2m5m5}
\eea
As explained in Ref.\cite{Ga1}, this solution can be thought of as 
corresponding to an M2-brane being stretched between two M5-branes:
when two M5-branes are brought together to intersect on a string,
one should think of the intersection as being a collapsed M2-brane.
Now we have to observe the following fact: if we start with this
configuration and we just take off one of the M5 branes, say the one 
oriented in (12345) spatial directions, we end with a configuration of an 
M2-brane (110) and an M5-brane (16789) intersecting on a string.
In fact, one can set $H_1 = 1$ in (\ref{m2m5m5}) to obtain
\beq
ds^{2} = H_{2}^{2/3}H_{3}^{1/3} \left[ H_2^{-1}H_3^{-1}(-dt^2 + ds^{2}) 
+ d{\vec{x}}^{2} + H_{2}^{-1} d{\vec{y}}^{2} + H_{3}^{-1} d w^{2} 
\right] ~,
\label{m2m5}
\eeq
which represents the configuration referred above that, in general, 
does not satisfy the harmonic superposition principle. In fact,
\beq
(H_{2}^{-1} \nabla_{x}^{2} + \nabla_{y}^{2}) H_{3} = 0 ~.
\label{infact}
\eeq
We will use a variant of our notation to call this configuration
$(M2|1|M5)$. It is interesting to mention that the intersection string
of this configuration is localized in the M5-brane but not in the M2-brane.
It has been studied earlier in Refs.\cite{Ts3,Ga1}. Note that a similar
solution with the intersection localized in the M2-brane instead of the
M5-brane exists \cite{Ts3} though it cannot be obtained within our
approach. Our construction of localized non-harmonic intersections as
$(M2|1|M5)$ is asymmetric from the very beginning, a fact reflected on
the different nature of $H_2$ and $H_3$. We would like to stress on the
fact that the amount of supersymmetry is not related to the degree of
localization of the membranes participating in a given configuration. 

~

There is a particular solution of this system with $H_3$
an harmonic function of the $x_a$'s coordinates\footnote{Also,
after this work was completed, an explicit solution of (\ref{infact})
was found for a D2-brane (also for a NS5-brane or a wave) localized
within a D6-brane, in the region close to the core of the D6-brane
\cite{ITY}.}. Thus, the configuration
$M2 \cap M5 (1)$ that depends on the overall transverse coordinates \cite{AEH} 
is a particular case of the (more general) solution $(M2|1|M5)$. One
can `deform' smoothly $(M2|1|M5)$ into $M2 \cap M5 (1)$, an operation that
cannot modify a relation between integers as it is the intersection rule
previously obtained.
It is then clear why we have obtained an intersection rule (\ref{dimint3})
that coincides with the one derived in Ref.\cite{AEH}. The same
reasoning can be straightforwardly applied to the whole set of
configurations we are considering in this Section.

\subsection{Configurations with NS and D branes}

As we discussed in section 3.2, the only configuration involving localized
NS-branes is $(1|NS5,NS5)$. Then, $\bar{q} = 1$, $p_t = 0$ and $q_3 = 1$. It
can be easily seen that equation (\ref{dimint3}) is not satisfied for a 
D1-brane but is precisely an identity for a fundamental string. That is,
one can accomodate a fundamental string along the common direction of the
solitonic NS-branes with vanishing binding energy. This configuration,
whose line element is given by
\[ ds^{2} = H_{1}^{3/4}H_{2}^{3/4}H_{3}^{1/4} \left[ 
H_1^{-1}H_2^{-1}H_3^{-1}(-dt^2 + ds^{2}) + H_1^{-1} d{\vec{x}}^{2} 
+ H_{2}^{-1} d{\vec{y}}^{2} \right] ~, \]
should be denoted by $(NS1|1|NS5,NS5)$, and it is clear that it can be
uplifted to eleven dimensions yielding (\ref{m2m5m5}).
One can follow the procedure mentioned in the previous subsection to extract one 
of the NS5-branes ending with the configuration $(NS1|1|NS5)$ first introduced 
in Ref.\cite{Ts3}.

~

Concerning localized intersection of NS-branes and D-branes, we have shown
in Section 3.2 that the only possible configurations that can be written in
terms of harmonic functions are $(q-3|NS5,Dq)$, for $3 \leq q \leq 8$. Then, 
for these configurations, one has $\bar{q} = q - 3$, $p_t = 1$ and
$q_3 = q - 2$. It is immediate to check in (\ref{dimint3}) that a 
D$q_3$-brane can be placed with vanishing binding energy as to build
the $(D(q-2)|q-3|NS5,Dq)$ configuration. It is worth to mention that, in
spite of the fact that the values $q=3$ and $q=7$ give enough room as to 
introduce a fundamental string and a NS5-brane, they do not satisfy the
intersection rule (\ref{dimint3}). Consequently, in our framework, we find
that only D-branes can be stretched between a NS5-brane and another D-brane.

~

At this point, we should come back to the examples that we started to discuss
on Section 3 which, after the addition of the third localized brane, are very
similar to the ones used in \cite{HW} for a $N=2$ configuration and in \cite{EGK} 
for a $N=1$ configuration.

~

Let us start with the NS5 (12345) - D5 (12789) configuration which preserves
1/4 of the supersymmetry. Then we have the intersection given by (12) and the
overall transverse coordinate given by (6). Then we see the possibility of
adding a third brane in the (126) direction and this will be a D3-brane.
This does not break further supersymmetries. The intersection
dimension agrees with formula (\ref{dimint3}). By studying the brane
dynamics and the conservation of magnetic charge, the appearance of the
D3-brane was explained in \cite{HW}. The difference between our case and
theirs is that our D3-brane is of infinite extension on $x_{6}$ direction
whereas their D3-brane is of finite extension on $x_{6}$ direction, this
extension being just the inverse of the coupling constant of the
gauge group $U(N)$ if we have N D3-branes on top of each other.
In our case, by having D3-branes with infinite extension on $x_{6}$ direction, 
would give only global groups, with coupling constant zero.

~

The second example is the one used in \cite{EGK}. Consider the localized
intersection of NS5 (12345) - D6 (123789). The intersection is given by (123) and the
overall transverse by (6). Then we see the possibility of introducing a 
D4-brane in the (1236) direction. Again our D4-brane is of infinite extension 
in the $x_{6}$ direction, as compared with \cite{EGK} where the D4-branes
are finite in that direction. Note, that we are not able to consider within our
framework the addition of the NS5'-branes that complete the configuration
considered in Ref.\cite{EGK}.

\subsection{Configurations with D-branes}

It is remarkable that, within our approach, there is a unique type IIB
configuration of three D-branes with vanishing binding energy,
$(D1|1|D5,D5)$, as can be easily seen from (\ref{dimint3}). There are,
of course, other solutions that can be constructed from it by 
delocalizing in a set of coordinates and applying T-dualities
along them \cite{Ts3}. Finally, it is interesting to mention that
the intersection rule (\ref{dimint3}) allows to locate a fundamental
string in the transverse direction of any of the configurations of
Section 3 but it does not allow the fundamental string to be in the
common direction of a pair of localized D-branes.

\section{Conclusions and Discussion}
In the present paper we discussed configurations with two and three branes
which are localized in their relative transverse coordinates. We obtained an
intersection rule for the case of two branes by solving the equations of
motion in a purely bosonic context. Our intersection rule
does not give new solutions and it just collects some of the solutions that
have appeared in the literature. It differs from the 
rule derived in Ref.\cite{AEH} --corresponding to intersecting branes which 
are localized in the overall transverse coordinates-- by a shift of two
units. We considered the introduction of a third brane with vanishing binding 
energy. We derived the corresponding intersection rule between the first two 
branes and the third one which happens to be identical to the one of 
Ref.\cite{AEH}. We clarified the origin of this coincidence by showing that 
the configurations which are localized in the overall transverse coordinates
can be obtained from those of three branes with localized intersections.
All branes are BPS states so for a threshold superposition we used the
`no-force' condition which led to strong simplifications of the
equations of motion. 

~

It would be interesting to generalize our work to include 
non-extreme\footnote{After the completion of this work, a paper
appeared covering this subject \cite{AIRV}.}
as well as extreme but non-supersymmetric configurations. 
In spite of the fact that it is not clear
if non-supersymmetric brane configurations represent consistent stable
backgrounds, an explicit construction of a generic configuration of
this kind could be a first step in order to study non-supersymmetric
gauge theories by using brane techniques. As well, it would be very
interesting to obtain within our approach other kind of brane configurations
which have been recently considered in the literature as p-branes at
angles \cite{BDL} and the so-called (p,q) polymers \cite{AH} or (p,q)
webs \cite{AHK} of branes. There are certain critical values for the angles 
of these configurations which might be thought of as being originated from 
the `no-force' conditions obtained from the Ricci tensor\footnote{We are
grateful to Amihay Hanany for his suggestions and comments on this
respect.}.
It should also be interesting to understand this kind of intersections
as the appearance of a certain soliton in the worldvolume field theory
of the complementary p-brane in the way recently introduced in 
Ref.\cite{To3,BGT,GGT}.

~

Another aspect that deserves a future study is the relation of the
kind of configurations appearing in our work and other geometries.
The intersecting brane configurations where all the functions depend on the overall
transverse coordinates have been related by T-dualities and changes of
coordinates with geometries of type $\mbox{adS}_{k}\times E^{l} \times S^{m}$
\cite{bpsk}. The main observation was that the harmonic functions lose the
constant part for a specific choice of transformations and so the geometry
is changed from a flat one to an $\mbox{adS}_{k}$ one. In 11 dimensions, they
have started from the general M2 (012)  - M5 (023456) solution and have considered 
the near horizon geometry for which the constant parts of the harmonic functions
become negligible. The spacetime factorizes as $\mbox{ad}S_{3}\times E^{5}
\times S^{3}$. In our case, we do not have a single radius variable and we could 
not identify a specific geometry when we neglect the constant term. Also, for
the three M-branes configuration $(M2|1|M5,M5)$, the function $H_{3}$ is not 
harmonic so we do not have the case of \cite{bpsk}.
We think that it would be very interesting to identify  the geometries that
are connected with our original ones by T-dualities and 
changes of coordinates of the near horizon geometry in 11 dimensions.
We hope to report on some of these problems elsewhere.

~

\section*{Acknowledgements}

We are pleased to thank Riccardo Argurio, Amihay Hanany, Laurent Houart and Arkady 
Tseytlin for their most interesting comments.
We would also like to thank Juan Maldacena for discussions and a critical reading 
of the manuscript. 
The work of J.D.E. is partially supported by a Spanish Ministry of Education and 
Culture fellowship.
Radu Tatar would like to thank Prof. Orlando Alvarez for advices.

\section*{Apendices}
\appendix
\section{Ricci Tensor of the Localized Intersections}
\setcounter{equation}{0}

In this Appendix we present the non-vanishing components of the Ricci
tensor that corresponds to the overlapping brane metric
\bea
ds^2 & = & -\,B^2 dt^2 + C^2 \delta_{ij} ds^i ds^j + X^2 \delta_{ab} 
dx^a dx^b + Y^2 \delta_{\a\b} dy^\a dy^\b \nonumber \\
& & +\,W^2 \delta_{\mu\nu} dw^{\mu} dw^{\nu} ~,
\label{overlap}
\eea
in order to clarify the origin of the `no-force' condition introduced
in this paper. They are:

\bea
R_{00} & = & Y^{-2}\left[\dery\ln{B}\cdot\dery\lnyal + 
\dery^2\ln{B}\right] \nonumber \\ & + & 
X^{-2}\left[\derx\ln{B}\cdot\derx\lnxal + \derx^2\ln{B}\right]
\label{roo}
\eea

\bea
R_{ij} & = & \delta_{ij} \left(Y^{-2}\left[\dery\ln{C}\cdot\dery\lnyal 
+ \dery^2\ln{C}\right] \right. \nonumber \\ & + & \left. 
X^{-2}\left[\derx\ln{C}\cdot\derx\lnxal + \derx^2\ln{C}\right] \right)
\label{rij}
\eea

\bea
R_{ab} & = & \delta_{ab} \left(Y^{-2}\left[\dery\ln{X}\cdot\dery\lnyal 
+ \dery^2\ln{X}\right] \right. \nonumber \\ & + & \left. 
X^{-2}\left[\derx\ln{X}\cdot\derx\lnxal + \derx^2\ln{X}\right] \right)
\nonumber \\ & + & X^{-2} \left(\derxa\derxb\lnxal + (p_1 - 
2)\derxa\ln{X}\derxb\ln{X} \right. \nonumber \\
& - & \left. 2\derxa\ln{X}\derxb\lnxal + p_2\derxa\ln{Y}\derxb\ln{Y} 
\right. \nonumber \\ & + & \left.\bar{q}\derxa\ln{C}\derxb\ln{C}
+ p_t\derxa\ln{W}\derxb\ln{W} + \derxa\ln{B}\derxb\ln{B} \right)
\label{rab}
\eea

\bea
R_{a\b} & = & - \derxa\ln{Y}\deryb\ln{BC^{\bar{q}}W^{p_t}X^{p_1-1}} 
- \derxa\ln{BC^{\bar{q}}W^{p_t}Y^{p_2-1}}\deryb\ln{X} \nonumber \\
& + & \bar{q}\derxa{C}\deryb{C} + \derxa{B}\deryb{B} + 
{p_t}\partial^a{W}\deryb{W} \nonumber \\
& - & \derxa\deryb\ln{X^{p_1-1}Y^{p_2-1}BC^{\bar{q}}W^{p_t}}
\label{rapb}
\eea

\bea
R_{\a\b} & = & \delta_{\a\b}
\left(Y^{-2}\left[\dery\ln{Y}\cdot\dery\lnyal 
+ \dery^2\ln{Y}\right] \right. \nonumber \\ & + & \left. 
X^{-2}\left[\derx\ln{Y}\cdot\derx\lnxal + \derx^2\ln{Y}\right] \right)
\nonumber \\ & + & Y^{-2} \left(\derya\deryb\lnyal + 
p_1\derya\ln{X}\deryb\ln{X} \right. \nonumber \\
& - & \left. 2\derya\ln{Y}\deryb\lnyal + (p_2 - 
2)\derya\ln{Y}\deryb\ln{Y} \right. \nonumber \\ & + & 
\left.\bar{q}\derya\ln{C}\deryb\ln{C}
+ p_t\derya\ln{W}\deryb\ln{W} + \derya\ln{B}\deryb\ln{B} \right)
\label{rpapb}
\eea

\bea
R_{\mu\nu} & = & \delta_{\mu\nu}
\left(Y^{-2}\left[\dery\ln{W}\cdot\dery\lnyal 
+ \dery^2\ln{W}\right] \right. \nonumber \\ & + & \left. 
X^{-2}\left[\derx\ln{W}\cdot\derx\lnxal + \derx^2\ln{W}\right] \right) ~.
\label{rmunu}
\eea

\noindent 
If we assume that all functions can be factorized 
\beq
{\cal F}(x^a,y^\a) \equiv {\cal F}_x(x^a){\cal F}_y(y^\a) ~,
\label{fact}
\eeq
and impose the `no-force' conditions,
\beq
B_xC_x^{\bar{q}}X_x^{p_1-2}Y_x^{p_2}W_x^{p_t} = 1 ~,
\label{aconst1}
\eeq
\beq
B_yC_y^{\bar{q}}X_y^{p_1}Y_y^{p_2-2}W_y^{p_t} = 1 ~.
\label{aconst2}
\eeq
these components get sensibly simplified as follows:

\beq
R_{00} = \Box\ln{B}
\label{ro}
\eeq

\beq
R_{ij} = \delta_{ij}\;\Box\ln{C}
\label{ri}
\eeq

\bea
R_{ab} & = & \delta_{ab}\;\Box\ln{X} + X^{-2} \left[ (p_1 - 
2)\derxa\ln{X}\derxb\ln{X} + p_2\derxa\ln{Y}\derxb\ln{Y} \right. \nonumber \\
& + & \left. \derxa\ln{B}\derxb\ln{B} + \bar{q}\derxa\ln{C}\derxb\ln{C} + 
p_t\derxa\ln{W}\derxb\ln{W} \right]
\label{ra}
\eea

\bea
R_{a\b} & = & (p_1 - 2)\derxa\ln{X}\deryb\ln{X} + (p_2 -
2)\derxa\ln{Y}\deryb\ln{Y} + \bar{q}\derxa{C}\deryb{C} \nonumber \\
& + & \derxa{B}\deryb{B} + {p_t}\derxa{W}\deryb{W} + 2 
\derxa\ln{Y}\deryb\ln{X}
\label{rpa}
\eea

\bea
R_{\a\b} & = & \delta_{\a\b}\;\Box\ln{Y} + Y^{-2} \left[ 
p_1\derya\ln{X}\deryb\ln{X} + (p_2 - 2)\derya\ln{Y}\deryb\ln{Y}
\right. \nonumber \\ & + & \left. \derya\ln{B}\deryb\ln{B} +
\bar{q}\derya\ln{C}\deryb\ln{C} + p_t\derya\ln{W}\deryb\ln{W} \right]
\label{rppa}
\eea

\beq
R_{\mu\nu} = \delta_{\mu\nu}\;\Box\ln{W} ~.
\label{rmu}
\eeq

\section{Third Brane with Zero Binding Energy}
\setcounter{equation}{0}
Let us now consider an additional contribution to all functions of the
metric
\bea
ds^2 & = & -\,\hat{B}^2 dt^2 + \hat{C}^2 \delta_{ij} ds^i ds^j + 
\hat{X}^2 \delta_{ab} dx^a dx^b + \hat{Y}^2 \delta_{\a\b} dy^\a dy^\b \nonumber \\
& & +\,\hat{W}^2 \delta_{\mu\nu} dw^{\mu} dw^{\nu} ~,
\label{overlap2}
\eea
in such a way that
\beq
\hat{\cal F}(x^a,y^\a) \equiv {\cal F}(x^a,y^\a) 
\tilde{\cal F}(x^a,y^\a) ~,
\label{fac2}
\eeq
where ${\cal F}$ are the solutions to the Einstein equations corresponding to
(\ref{ro})--(\ref{rmu}). The modified Ricci tensor turns out to be
\beq
\hat{R}_{MN} = \tX^{-2}R_{MN} + \tR_{MN} ~,
\label{modricci}
\eeq
where
\bea
\tR_{00} & = & \hat{Y}^{-2}\left[\dery\ln{\hat{B}}\cdot\dery\lnyallt + 
\dery^2\ln{\tB}\right] \nonumber \\ & + & 
\hat{X}^{-2}\left[\derx\ln{\hat{B}}\cdot\derx\lnxallt + \derx^2\ln{\tB}\right]
\label{roou}
\eea

\bea
\tR_{ij} & = & \delta_{ij} \left(\hat{Y}^{-2}\left[\dery\ln{\hat{C}}\cdot\dery\lnyallt 
+ \dery^2\ln{\tC}\right] \right. \nonumber \\ & + & \left. 
\hat{X}^{-2}\left[\derx\ln{\hat{C}}\cdot\derx\lnxallt + \derx^2\ln{\tC}\right] 
\right)
\label{riju}
\eea

\bea
\tR_{ab} & = & \delta_{ab} \left(\hat{Y}^{-2}\left[\dery\ln{\hat{X}}\cdot\dery\lnyallt 
+ \dery^2\ln{\tX}\right] \right. \nonumber \\ & + & \left. 
\hat{X}^{-2}\left[\derx\ln{\hat{X}}\cdot\derx\lnxallt + \derx^2\ln{\tX}\right] \right)
\nonumber \\ & + & \hat{X}^{-2} \left(\derxa\derxb\lnxallt + (p_1 - 
2)\derxa\ln{\hat{X}}\derxb\ln{\hat{X}} \right. \nonumber \\
& - & \left. 2\derxa\ln{\hat{X}}\derxb\lnxallt + 
p_2\derxa\ln{\hat{Y}}\derxb\ln{\hat{Y}} \right. \nonumber \\ 
& + & \left.\bar{q}\derxa\ln{\hat{C}}\derxb\ln{\hat{C}}
+ p_t\derxa\ln{\hat{W}}\derxb\ln{\hat{W}} + \derxa\ln{\hat{B}}\derxb\ln{\hat{B}} \right.
\nonumber \\ & - & \left. (p_1 - 2)\derxa\ln{X}\derxb\ln{X} - p_2\derxa\ln{Y}\derxb\ln{Y}
- \bar{q}\derxa\ln{C}\derxb\ln{C} \right. \nonumber \\ 
& - & \left. p_t\derxa\ln{W}\derxb\ln{W} - \derxa\ln{B}\derxb\ln{B} \right)
\label{rabu}
\eea

\bea
\tR_{a\b} & = & - 
\derxa\ln{\hat{Y}}\deryb\ln{\hat{B}\hat{C}^{\bar{q}}\hat{W}^{p_t}\hat{X}^{p_1-1}} 
- \derxa\ln{\hat{B}\hat{C}^{\bar{q}}\hat{W}^{p_t}\hat{Y}^{p_2-1}}\deryb\ln{\hat{X}} 
\nonumber \\ & + & \bar{q}\derxa\ln{\hat{C}}\deryb\ln{\hat{C}} + 
\derxa\ln{\hat{B}}\deryb\ln{\hat{B}} + {p_t}\derxa\ln{\hat{W}}\deryb\ln{\hat{W}}
\nonumber \\ & - &
\derxa\deryb\ln{\hat{X}^{p_1-1}\hat{Y}^{p_2-1}\hat{B}\hat{C}^{\bar{q}}\hat{W}^{p_t}}
- \bar{q}\derxa\ln{C}\deryb\ln{C} \nonumber \\
& - & \derxa\ln{B}\deryb\ln{B} - {p_t}\derxa\ln{W}\deryb\ln{W}
\label{rapbu}
\eea

\bea
\tR_{\a\b} & = & \delta_{\a\b}
\left(\hat{Y}^{-2}\left[\dery\ln{\hat{Y}}\cdot\dery\lnyallt 
+ \dery^2\ln{\tY}\right] \right. \nonumber \\ & + & \left. 
\hat{X}^{-2}\left[\derx\ln{\hat{Y}}\cdot\derx\lnxallt + \derx^2\ln{\tY}\right] \right)
\nonumber \\ & + & \hat{Y}^{-2} \left(\derya\deryb\lnyallt + 
p_1\derya\ln{\hat{X}}\deryb\ln{\hat{X}} \right. \nonumber \\
& - & \left. 2\derya\ln{\hat{Y}}\deryb\lnyallt + (p_2 - 
2)\derya\ln{\hat{Y}}\deryb\ln{\hat{Y}} \right. \nonumber \\ & + & 
\left.\bar{q}\derya\ln{\hat{C}}\deryb\ln{\hat{C}} + 
p_t\derya\ln{\hat{W}}\deryb\ln{\hat{W}} + \derya\ln{\hat{B}}\deryb\ln{\hat{B}} \right.
\nonumber \\ & - & 
\left. p_1\derya\ln{X}\deryb\ln{X} - (p_2 - 2)\derya\ln{Y}\deryb\ln{Y} -
\bar{q}\derya\ln{C}\deryb\ln{C} \right. \nonumber \\
& - & \left. p_t\derya\ln{W}\deryb\ln{W} - \derya\ln{B}\deryb\ln{B} \right)
\label{rpapbu}
\eea

\bea
\tR_{\mu\nu} & = & \delta_{\mu\nu}
\left(\hat{Y}^{-2}\left[\dery\ln{\hat{W}}\cdot\dery\lnyallt 
+ \dery^2\ln{\tW}\right] \right. \nonumber \\ & + & \left. 
\hat{X}^{-2}\left[\derx\ln{\hat{W}}\cdot\derx\lnxallt + \derx^2\ln{\tW}\right]
 \right) ~.
\label{rmunuu}
\eea

These expressions suggest that the `no-force' condition that must be satisfied in 
order that the third brane could be bounded to the old configuration with zero
binding energy are:
\beq
\tB\tC^{\bar{q}}\tX^{p_1-2}\tY^{p_2}\tW^{p_t} = 1 ~,
\label{aconst3}
\eeq
\beq
\tB\tC^{\bar{q}}\tX^{p_1}\tY^{p_2-2}\tW^{p_t} = 1 ~.
\label{aconst4}
\eeq
The Ricci tensor gets simplified after (\ref{aconst3}) and (\ref{aconst4}) to

\beq
\tR_{00} = \tBox\ln{\tB}
\label{rooz}
\eeq

\beq
\tR_{ij} = \delta_{ij}\;\tBox\ln{\tC}
\label{rijz}
\eeq

\bea
\tR_{ab} & = & \delta_{ab}\;\tBox\ln{\tX} + \hat{X}^{-2} \left( (p_1 - 
2)\derxa\ln{\hat{X}}\derxb\ln{\hat{X}} + p_2\derxa\ln{\hat{Y}}\derxb\ln{\hat{Y}}
\right. \nonumber \\
& + & \left. \bar{q}\derxa\ln{\hat{C}}\derxb\ln{\hat{C}} + 
p_t\derxa\ln{\hat{W}}\derxb\ln{\hat{W}} + \derxa\ln{\hat{B}}\derxb\ln{\hat{B}}
\right. \nonumber \\
& - & \left. (p_1 - 2)\derxa\ln{X}\derxb\ln{X} 
- p_2\derxa\ln{Y}\derxb\ln{Y} - \bar{q}\derxa\ln{C}\derxb\ln{C} 
\right. \nonumber \\ 
& - & \left. p_t\derxa\ln{W}\derxb\ln{W} 
- \derxa\ln{B}\derxb\ln{B} \right)
\label{rabz}
\eea

\bea
R_{a\b} & = & (p_1 - 2)\derxa\ln{\hat{X}}\deryb\ln{\hat{X}} + (p_2 -
2)\derxa\ln{\hat{Y}}\deryb\ln{\hat{Y}} + \bar{q}\derxa{\hat{C}}\deryb{\hat{C}} 
\nonumber \\
& + & \derxa{\hat{B}}\deryb{\hat{B}} + {p_t}\derxa{\hat{W}}\deryb{\hat{W}} + 2 
\derxa\ln{\hat{Y}}\deryb\ln{\hat{X}} - (\mbox{unhatted})
\label{rpabz}
\eea

\bea
\tR_{\a\b} & = & \delta_{\a\b}\;\tBox\ln{\tY} + \hat{Y}^{-2} 
\left(p_1\derya\ln{\hat{X}}\deryb\ln{\hat{X}} + (p_2 - 
2)\derya\ln{\hat{Y}}\deryb\ln{\hat{Y}} \right. \nonumber \\ 
& + & \left. \bar{q}\derya\ln{\hat{C}}\deryb\ln{\hat{C}} + 
p_t\derya\ln{\hat{W}}\deryb\ln{\hat{W}} + \derya\ln{\hat{B}}\deryb\ln{\hat{B}} 
\right. \nonumber \\ & - & 
\left. p_1\derya\ln{X}\deryb\ln{X} - (p_2 - 2)\derya\ln{Y}\deryb\ln{Y} -
\bar{q}\derya\ln{C}\deryb\ln{C} \right. \nonumber \\
& - & \left. p_t\derya\ln{W}\deryb\ln{W} - \derya\ln{B}\deryb\ln{B} \right)
\label{rpapbz}
\eea

\beq
\tR_{\mu\nu} = \delta_{\mu\nu}\;\tBox\ln{\tW}.
\label{rmunuz}
\eeq

\end{document}